\documentclass[12pt]{article}
\usepackage{amsmath}
\usepackage{graphicx} 
\usepackage{amsmath,amssymb,epsfig,bm}
\newcommand\x{\mathbf{x}}
\newcommand\<{\langle}
\renewcommand\>{\rangle}
\renewcommand\d{\partial}

\renewcommand\d{\partial}

\begin{document}

\title{{\bf
Viscosity in Strongly Interacting Quantum Field Theories 
from Black Hole Physics}}
\author{P.\,K.~Kovtun,$^1$\footnote{email: {\tt kovtun@kitp.ucsb.edu}} $~$
D.\,T.~Son,$^2$\footnote{email: {\tt son@phys.washington.edu}} \,
and A.\,O.~Starinets$^3$\footnote{email: {\tt starina@perimeterinstitute.ca}}
\vspace{6pt}
\\
{\small\em $^1$Kavli Institute for Theoretical Physics, University 
of California, Santa Barbara, CA 93106, USA}\vspace{-6pt}\\
{\small\em $^2$Institute for Nuclear Theory, University of Washington,
Seattle, WA 98195-1550, USA}\vspace{-6pt}\\
{\small\em $^3$Perimeter Institute for Theoretical Physics,
Waterloo, Ontario N2L 2Y5, Canada}
}
\date{December 2004}
\maketitle

\begin{abstract}
The ratio of shear viscosity to volume density of entropy can be used to
characterize how close a given fluid is to being perfect.  Using
string theory methods, we show that this ratio is equal to a universal
value of $\hbar/4\pi k_B$ for a large class of strongly interacting
quantum field theories whose dual description involves black holes in
anti--de Sitter space.  We provide evidence that this value may serve
as a lower bound for a wide class of systems, thus suggesting
that black hole horizons are dual to the most ideal fluids.
\end{abstract}
\begin{flushleft} 
{\small PACS numbers: 11.10.Wx, 04.70.Dy, 11.25.Tq, 47.75.+f}
\end{flushleft}
\vspace{-6in}

\newpage

\emph{Introduction.}---It has been known since the discovery of Hawking
radiation~\cite{Hawking:sw} that black holes are endowed with
thermodynamic properties such as entropy and temperature, as first
suggested by Bekenstein \cite{Bekenstein:ur} based on the analogy
between black hole physics and equilibrium thermodynamics.
In higher-dimensional gravity theories there exist solutions called
black branes, which are black holes with translationally invariant
horizons \cite{Horowitz:1991cd}.  For these solutions, thermodynamics
can be extended to \emph{hydrodynamics}---the theory that describes
long-wavelength deviations from thermal equilibrium \cite{landau}.  
In addition to
thermodynamic properties such as temperature and entropy, black branes
possess hydrodynamic characteristics of continuous fluids: viscosity,
diffusion constants, etc.  From the perspective of the holographic
principle \cite{'tHooft:gx, Susskind:1994vu}, a black brane corresponds
to a certain finite-temperature quantum field theory in fewer
number of spacetime dimensions, and the hydrodynamic behavior of a
black-brane horizon is identified with the hydrodynamic behavior of
the dual theory.
For these field theories, in this Letter we show that
the ratio of the shear viscosity to the volume density of entropy has
a universal value
\begin{equation}
  \frac\eta s = \frac\hbar{4\pi k_B} 
  \approx6.08\times10^{-13}~\textrm{K} \,\textrm{s}\, .
\end{equation}
Furthermore, we shall argue that this is the lowest bound on the ratio
$\eta/s$ for a wide class of thermal quantum field theories.
\vspace{6pt}

\emph{Viscosity and graviton absorption.}---%
Consider a thermal field theory whose dual holographic description involves 
a $D$-dimensional 
black-brane metric of the form
\begin{equation}\label{metric_g}
\begin{split}
  d s^2 &= g^{(0)}_{MN} dx^M dx^N  
  = f (\xi)\,  (d x^2 + d y^2) + 
  g_{\mu\nu} (\xi) \, d \xi^{\mu} d \xi^{\nu}\,.
\end{split}
\end{equation}
[The $O(2)$ symmetry of the background is required for the 
existence of the shear hydrodynamic mode in the dual theory, thus 
making the notion of shear viscosity meaningful.]
One can have in
mind, as an example, the near-extremal D3-brane in type IIB
supergravity, dual to finite-temperature ${\cal N}=4$ supersymmetric
SU($N_c$) Yang--Mills theory in the limit of large $N_c$ and large 't Hooft
coupling \cite{Maldacena:1997re,Gubser:1998bc,Witten:1998qj,Aharony:1999ti},
\begin{equation}\label{D3}
\begin{split}
  ds^2 &= \frac{r^2}{R^2}\Biggl[ - \left( 1-\frac{r_0^4}{r^4} \right)\,
          d t^2  + d x^2 + d y^2 + d z^2\Biggr] 
  + \frac{R^2}{r^2 (1 - r_0^4/r^4)} d r^2\,,
\end{split}
\end{equation}
but our discussion will be quite general.
All black branes have an event horizon [$r=r_0$ for the
metric~(\ref{D3})], which is extended along several spatial dimensions
[$x$, $y$, $z$ in the case of~(\ref{D3})].  The dual field theory is
at a finite temperature, equal to the Hawking temperature of the black
brane.

The entropy of the dual field theory is equal to the entropy of the
black brane, which is proportional to the area of its event horizon,
\begin{equation}
  S = \frac A{4G}\,,
\end{equation}
where $G$ is Newton's constant (we set $\hbar=c=k_B=1$).  For black
branes $A$ contains a trivial infinite factor $V$ equal to the spatial
volume along directions parallel to the horizon.  The entropy density
$s$ is equal to $a/(4G)$, where $a = A/V$.


The shear viscosity of the dual theory can be computed from gravity 
in a number of equivalent approaches \cite{Policastro:2001yc, 
Policastro:2002se, Kovtun:2003wp}.
Here we use Kubo's formula, which relates viscosity
to equilibrium correlation functions. In a rotationally invariant field 
theory,
\begin{equation}\label{Kubo}
   \eta = \lim_{\omega\to0} \frac1{2\omega}\int\!dt\,d\x\,
   e^{i\omega t} \,\< [T_{xy}(t,\x),\, T_{xy}(0,{\bf 0})] \>\,.
\end{equation}
Here $T_{xy}$ is the $xy$ component of the stress-energy tensor (one
can replace $T_{xy}$ by any component of the traceless part of the
stress tensor).  We shall now relate the right-hand side
of~(\ref{Kubo}) to the absorption cross section of
low-energy gravitons.

According to the gauge--gravity duality \cite{Aharony:1999ti}, 
the stress-energy tensor
$T_{\mu\nu}$ couples to metric perturbations at the boundary.  
Following Klebanov \cite{Klebanov:1997kc,Gubser:1997yh}, let us
consider a graviton of frequency $\omega$, polarized in the $xy$
direction, and propagating perpendicularly to the brane.  In the field theory
picture, the absorption cross section of the graviton by the brane 
measures the imaginary part of
the retarded Greens function of the operator coupled to $h_{xy}$,
i.e., $T_{xy}$,
\begin{equation}\label{sigma-abs}
\begin{split}
  \sigma_{\rm abs}(\omega) &= -\frac{2\kappa^2}\omega 
   \,\textrm{Im}\, G^{\rm R}(\omega) 
  = \frac{\kappa^2}\omega\int\!dt\,d\x\, e^{i\omega t}\,
   \< [ T_{xy}(t,\x),\, T_{xy}(0,{\bf 0})]\>\,,
\end{split}
\end{equation}
where $\kappa = \sqrt{8\pi G}$ appears due to the normalization of the
graviton's action.  Comparing (\ref{Kubo}) and (\ref{sigma-abs}), we
find
\begin{equation}\label{eta-abs}
  \eta = \frac{\sigma_{\rm abs}(0)}{2\kappa^2} =
         \frac{\sigma_{\rm abs}(0)}{16\pi G} \ .
\end{equation}\vspace{0pt}

\emph{Graviton absoprtion cross section at low energy.}---%
The absorption cross section $\sigma_{\rm abs}$ is calculable
classically by solving the linearised wave equation for $h^x_y$.  
We now show that $h^x_y=h_{xy}/f$
obeys the equation for a minimally coupled massless scalar in the
background~(\ref{metric_g}).
This is similar to
cosmological tensor perturbations on a Friedmann--Robertson--Walker
spacetime, which obey the equation for a massless scalar
field~\cite{Liddle:2000cg}.

Consider small perturbations around the metric,
$g_{MN}= g_{MN}^{(0)} + h_{MN}$.  We assume that the only
non-vanishing component of $h_{MN}$ is $h_{xy}$, and that it does
not depend on $x$ and $y$: $h_{xy}=h_{xy}(\xi)$.  This field has spin 2
under the $O$(2) rotational symmetry in the $xy$ plane, which implies
that all other components of $h_{MN}$ can be consistently set to
zero \cite{Policastro:2002se}.
Einstein's equations can be written in the form
\begin{equation}
R_{M N} = T_{M N} - \frac T{D-2} \, g_{M N}\,,
\label{es_eq}
\end{equation}
where the stress-energy tensor $T_{MN}$ and its trace $T$ depend on
other fields such as the dilaton and various forms supporting the
background (\ref{metric_g}), for example, the fields appearing in the
low-energy type II string theory.  
Again, $O$(2) $xy$ rotational
symmetry implies that all perturbations of matter fields can be 
set to zero consistently.  
Thus when $M$ and $N$ are $x$ or $y$, the right-hand 
side of Einstein's equations reads ($\alpha, \beta = x, y$)
\begin{equation}
   T_{\alpha\beta} - \frac T{D-2} \, g_{\alpha\beta} = 
   -\left({\cal L} + \frac{T^{(0)}}{D-2}\right) 
   (\delta_{\alpha\beta}f + h_{\alpha\beta})\,,
\end{equation}
where ${\cal L}$ is the Lagrange density of matter fields 
and $T^{(0)}$ is the trace of
the unperturbed stress-energy tensor.
\begin{equation}
T_{MN} = - g_{MN} \, {\cal L} +\cdots\,,
\label{stress}
\end{equation}
where ${\cal L}$ represents the Lagrangian density of the fields,
and dots denote terms of second and higher orders 
in the perturbation $h_{xy}$.  
Substituting  the unperturbed
metric~(\ref{metric_g}) into Einstein's equations, one finds
\begin{equation}\label{Einstein-unpert}
  \frac12 \left[  \frac{\Box f}f - \frac{(\d f)^2}{f^2}\right]
  = {\cal L} + \frac{T^{(0)}}{D-2}\,.
\end{equation}

Expanding Einstein's equations to linear order in $h_{xy}$, one has
\begin{equation}\label{Einstein-pert}
\begin{split}
  R_{x y} &= - \frac12 \Box h_{xy} + 
   \frac1f \d^\mu\! f\, \partial_{\mu} h_{xy} 
   - \frac{(\d f)^2}{2f^2} h_{xy} 
  =
   -\left({\cal L} + \frac{T^{(0)}}{D-2}\right)h_{xy}\,.
\end{split}
\end{equation}
Combining Eqs.~(\ref{Einstein-unpert}) and (\ref{Einstein-pert}), we
obtain an equation for $h_{xy}$:
\begin{equation}
  \Box \, h_{xy} - 2\, \frac{\d^\mu\!f}f\, \d_{\mu} h_{xy}  + 
  2 \, \frac{(\d f)^2}{f^2} \, h_{xy} - \frac{\Box f}f \, h_{xy} = 0\,.
\label{minscal}
\end{equation}
Changing the variable from $h_{xy}$ to $h^x_y=h_{xy}/f$, one can see
that $h^x_y$ indeed satisfies the equation for a minimally coupled massless 
scalar: $\Box h^x_y = 0$.  
The absorption cross section
of a graviton is therefore the same as that of the scalar.

The absorption cross section for the scalar is constrained by
a theorem \cite{Das:1996we,Emparan:1997iv}, which states that in the
low-frequency limit $\omega\to0$ this cross section is equal to the
area of the horizon, $\sigma_{\rm abs}=a$.  Since $s=a/4G$, one
immediately finds that
\begin{equation}
  \frac\eta s = \frac\hbar{4\pi k_B}\,,
\label{the_ratio}
\end{equation}
where $\hbar$ and $k_B$ are now restored.  This shows that the ratio
$\eta/s$ does not depend on the concrete form of the metric within the
assumptions of Refs. \cite{Das:1996we,Emparan:1997iv}.  Indeed, this ratio is
the same for D$p$- (\cite{Policastro:2001yc,Kovtun:2003wp}), M2- and
M5- (\cite{Herzog:2002fn}) branes and for deformations of the D3
metric \cite{Kovtun:2003wp,Buchel:2003tz}.  This fact is very
surprising, given that the corresponding dual field theories are
very different.  We do not have an explanation for the constancy of
$\eta/s$ in these theories based on field-theoretical arguments alone.
\vspace{6pt}

\emph{A viscosity bound conjecture.}---%
Most quantum field theories do not have simple gravity duals.  Is our
result relevant in a broader setting?  We speculate that the ratio
$\eta/s$ has a lower bound
\begin{equation}\label{bound}
  \frac\eta s \geqslant \frac\hbar{4\pi k_B}
\end{equation}
for all relativistic quantum field theories at finite temperature and
zero chemical potential.  The inequality is saturated by theories with
gravity duals.

One argument supporting the bound (\ref{bound}) comes from the
Heisenberg uncertainty principle.  The viscosity of a plasma is
proportional to $\epsilon\tau_{\rm mft}$, where $\epsilon$ is the
energy density and $\tau_{\rm mft}$ is the typical mean free time of a
quasiparticle.  The entropy density, on the other hand, is
proportional to the density of quasiparticles, $s\sim k_B n$.
Therefore $\eta/s \sim k_B^{-1}\tau_{\rm mft} \epsilon/n$.  Now
$\epsilon/n$ is the average energy per particle.  According to the
uncertainty principle, the product of the energy of a quasiparticle
$\epsilon/n$ and its mean free time $\tau_{\rm mft}$ cannot be smaller
than $\hbar$, otherwise the quasiparticle concept does not make sense.
Therefore we obtain, from the uncertainty principle alone, that
$\eta/s\gtrsim \hbar/k_B$, which is (\ref{bound}) without the
numerical coefficient of $1/(4\pi)$.  We also conclude that $\eta/s$
is much larger than $\hbar/k_B$ in weakly coupled theories (where the
mean free time is large).

Another piece of evidence supporting the bound (\ref{bound}) comes
from a recent calculation \cite{Buchel:2004di} of $\eta/s$ in the
${\cal N}=4$ supersymmetric $SU(N_c)$ Yang--Mills theories in the regime of
infinite $N_c$ and large, but finite, 't Hooft coupling $g^2N_c$.  The
first correction in inverse powers of $g^2N_c$ corresponds to the
first string theory correction to Einstein's gravity.  The result reads
\begin{equation}
\frac{\eta}{s} = \frac\hbar{4 \pi k_B} \left( 1 + 
  \frac {135\zeta(3)}{8 (2 g^2 N_c)^{3/2}} + \cdots \right)
\end{equation}
where $\zeta (3)\approx 1.2020569...$ is Ap\'{e}ry's constant.
The correction is positive, in accordance with (\ref{bound}).  It is
natural to assume that $\eta/s$ is larger than the bound for all
values of the 't~Hooft coupling (Fig.~1).
\begin{figure}[htb]
\centerline{\includegraphics[width=12cm,angle=0]{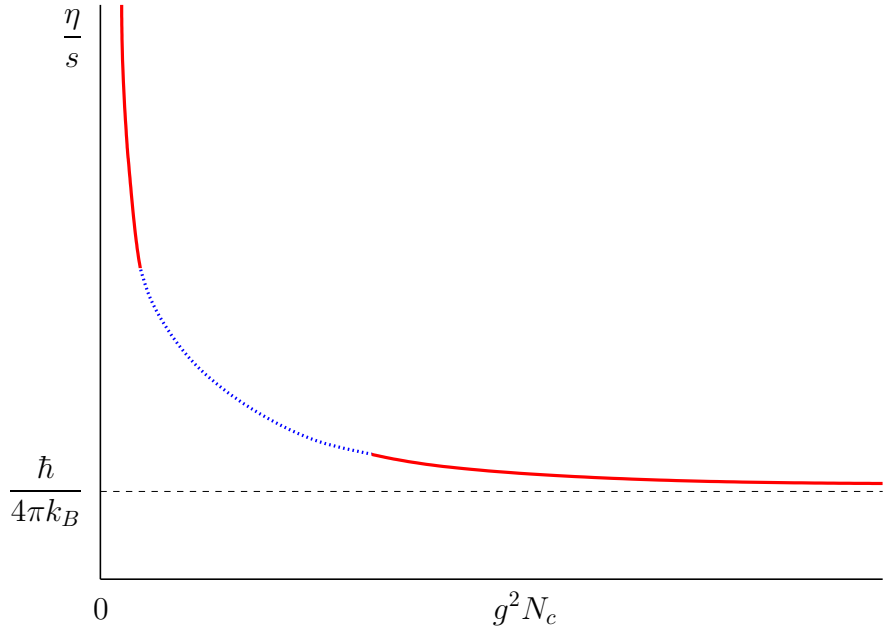}}
\caption{\label{fig1} The dependence 
of the ratio $\eta/s$ on the 't Hooft coupling
$g^2N_c$ in ${\cal N}=4$ supersymmetric Yang--Mills theory.  The ratio
diverges in the limit $g^2 N_c\to0$ and approaches 
$\hbar/{4 \pi k_B}$
from above as $g^2N_c\to\infty$.  
The ratio is unknown in the regime
of intermediate 
't~Hooft coupling.
}
\end{figure}

The bound (\ref{bound}), in contrast to the entropy
bound \cite{Bousso:1999xy} and Bekenstein's
bound \cite{Bekenstein:jp}, does not involve the speed of light $c$ and
hence is nontrivial when applied to nonrelativistic systems.  However,
the range of applicability of (\ref{bound}) to nonrelativistic systems
is less certain.  On the one hand, by subdividing the molecules of a gas
to an ever-increasing number of non-identical species one can increase
the entropy density (by adding the Gibbs mixing entropy) without
substantially affecting the viscosity.  On the other hand, the
conjectured bound is far below the ratio of $\eta/s$ in any laboratory
liquid.  For water under normal conditions $\eta/s$ is 380 times
larger than $\hbar/(4\pi k_B)$.  Using standard
tables \cite{webbook,codata} one can find $\eta/s$ for many liquids and
gases at different temperatures and pressures.  Figure 2 shows
temperature dependence of $\eta/s$, normalized by $\hbar/(4\pi k_B)$,
for a few substances at different pressures.  It is clear that the
viscosity bound is well satisfied for these substances.  Liquid helium
reaches the smallest value of $\eta/s$, but this value still exceeds
the bound by a factor of about 9.  We speculate that the bound
(\ref{bound}) is valid at least for a 
single-component non-relativisitic gas of
particles with spin 0 or 1/2.
\begin{figure}[htb]
\centerline{\includegraphics[width=15cm,angle=0]{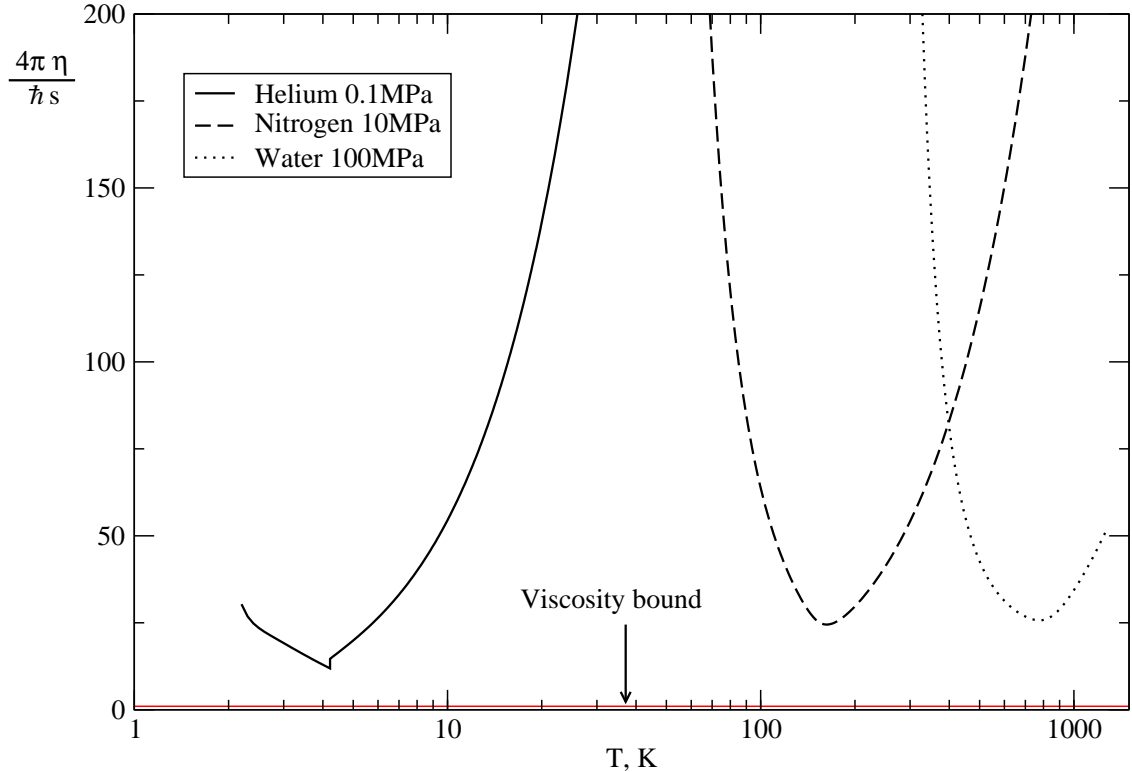}
}
\caption{\label{fig2}The viscosity-entropy ratio for some common substances:
helium, nitrogen and water.  The ratio is always substantially larger
than its value in theories with gravity duals, represented by the
horizontal line marked ``viscosity bound.''}
\end{figure}
\vspace{6pt}

\emph{Discussion.}---%
It is important to avoid some common misconceptions which at first
sight seem to invalidate the viscosity bound.  
Somewhat counterintuitively, 
a near-ideal gas has a very large viscosity due to the
large mean free path.  Likewise, superfluids have finite and
measurable shear viscosity associated with the normal component,
according to Landau's two-component theory.

The bound~(\ref{bound}) is most useful for strongly interacting
systems where reliable theoretical estimates of the viscosity are 
not available.
One of such systems is the quark-gluon plasma (QGP) created in heavy
ion collisions which behaves in many respects as a droplet of a
liquid.  
There are experimental hints
that the viscosity of the QGP at temperatures achieved by the
Relativistic Heavy Ion Collider is surprisingly small, possibly close
to saturating the viscosity bound \cite{Shuryak:2003xe}.  
Another possible application of the viscosity bound is trapped atomic
gases.  By using the Feshbach resonance, strongly interacting Fermi
gases of atoms have been created recently.  These gases have been
observed to behave hydrodynamically \cite{OHara} and should have finite
shear viscosity at nonzero temperature.
It would be very interesting to test experimentally 
whether the shear viscosity of these gases satisfies the 
conjectured 
 bound.



\vspace{6pt}

This work was supported by DOE grant DE-FG02-00ER41132, the
National Science Foundation and the Alfred P.~Sloan Foundation.




\end{document}